\documentclass[aps,prl,twocolumn,showpacs,groupedaddress,amssymb]{revtex4}

\usepackage{graphicx} 

\begin{document}
\title{Microstructure of a liquid complex (dusty) plasma under shear}

\author{V. Nosenko}
\email{nosenko@mpe.mpg.de}
\author{A. V. Ivlev}
\author{G. E. Morfill}
\affiliation{Max-Planck-Institut f\"{u}r extraterrestrische Physik, D-85741 Garching, Germany}

\date{\today}
\begin{abstract}
The microstructure of a strongly coupled liquid undergoing a shear flow was studied experimentally. The liquid was a shear melted two-dimensional plasma crystal, i.e., a single-layer suspension of micrometer-size particles in a rf discharge plasma. Trajectories of particles were measured using video microscopy. The resulting microstructure was anisotropic, with compressional and extensional axes at around $\pm 45^{\circ}$ to the flow direction. Corresponding ellipticity of the pair correlation function $g({\bf r})$ or static structure factor $S(\bf{k})$ gives the (normalized) shear rate of the flow.
\end{abstract}
 \maketitle

Shear flows are ubiquitous components of almost any flow in liquids. They are important in many industrial applications and pose interesting questions in fluid mechanics. One question, which goes beyond the conventional hydrodynamic approach (based on the Navier-Stokes equation) is a relation between the strain rate in a flowing liquid and its microstructure. It has long been recognized that the microstructure of various sheared liquids is distorted in a way similar to an elastically deformed solid (glass) \cite{Clark:80,Hanley:83,Henrich:09,Fuchs:09,Brader:10}. In fact, the very existence of shear viscosity in a liquid is closely related to this distortion. Experimentally, this distortion can be indirectly seen in the static structure factor of a sheared liquid deduced from scattering measurements. Direct observation of the microstructure distortion in real space is impractical for regular liquids and is only possible in simulations or model systems where the motion of individual particles -- proxy ``atoms'' -- can be resolved. Suitable model systems are colloidal dispersions \cite{Clark:80,Hanley:83,Fuchs:09,Brader:10} and complex (dusty) plasmas \cite{Morfill:09,ShuklaBook}.

Complex plasma is a suspension of fine solid particles in a weakly ionized gas \cite{Morfill:09,ShuklaBook}. Micron-size particles acquire high electric charge (usually negative) and may self-organize in ordered structures. The particle dynamics is governed by interparticle interactions as well as by interactions with the surrounding plasma. The latter can lead to various instabilities resulting in anomalous ``heating.'' However, for single-layer suspensions these instabilities are well understood and can be easily suppressed by proper choice of experimental parameters \cite{Couedel:11}. In this case the particle kinetic temperature is a well-defined value determined by the temperature of ambient gas. Depending on the particle coupling strength (i.e, the ratio of the pair interaction energy to temperature), a complex plasma can be in a liquid or solid state. The motion of individual particles is fully resolved in real time, allowing direct observation of their dynamics at the ``atomistic'' level. In two-dimensional (2D) complex plasmas, particles are arranged in a single layer and are therefore easy to observe using video microscopy.

Complex plasmas belong to a broad class of soft condensed matter: Typical shear modulus of a crystalline 2D sample is around $10^{-13}$~N/mm \cite{Nosenko:11PRL_disl}, so that they can be easily manipulated, e.g., by applying radiation pressure of a laser beam \cite{Chan:04,Nosenko:09PRL}. This allows one to study a broad range of phenomena from nucleation and motion of single dislocations \cite{Nosenko:11PRL_disl} to shear flows \cite{Nosenko:04PRL_visc}. In Ref.~\cite{Nosenko:04PRL_visc}, the shear viscosity of liquid complex plasma was calculated using a fit of measured particle velocity profiles to a Navier-Stokes model. However, any changes occurring in the microstructure of sheared liquid complex plasma have not been studied so far.

In this paper, we experimentally study the microstructure of a sheared 2D complex plasma at various levels of strain rate. We find that the microstructure is distorted, with compressional and extensional axes at around $\pm 45^{\circ}$ to the flow direction. The magnitude of distortion is defined by the (normalized) shear flow's strain rate.

\begin{figure}
\centering
\includegraphics[width=\columnwidth]{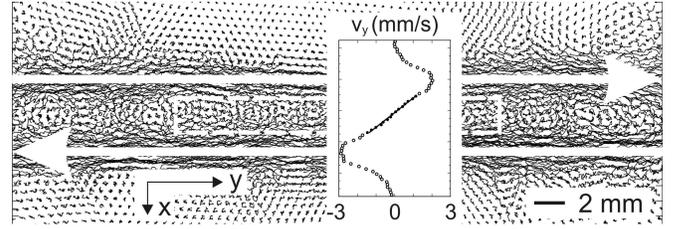}
\caption {\label {exp_layout} Shear flow in a planar Couette configuration in a strongly coupled 2D complex plasma is sustained by a pair of counterpropagating laser beams (indicated by arrows). Particle trajectories during $0.83$~s of a steady-state flow are shown. The dashed-line box defines the region of interest where the analysis of liquid microstructure was performed. The inset shows the transverse profile of particle velocity $v_y(x)$. The manipulation laser power was $P_{\rm laser}=1.25$~W.}
\end{figure}

Our experimental setup was a modified GEC (Gaseous Electronics Conference) rf reference cell \cite{Nosenko:11PRL_disl}. Plasma was produced using a capacitively coupled rf discharge in argon at $0.66$~Pa. A single layer of dust particles was suspended in the plasma sheath of the lower rf electrode. The microspheres made of melamine formaldehyde had a diameter of $9.19\pm0.09$~$\mu$m, a mass $m=6.15 \times 10^{-13}$~kg, and acquired an electric charge of $Q=-17~000\pm1700e$. The suspension included around $8000$ particles and had a diameter of $\approx60$~mm, the mean interparticle distance in the center was $\Delta=0.55$~mm [measured from the first peak of the pair correlation function $g(r)$]. The neutral gas damping rate was $\nu=0.77~{\rm s}^{-1}$.

The like-charged particles repelled each other via a screened Coulomb, or Yukawa potential: $U(r)=Q(4\pi\epsilon_0r)^{-1}{\rm exp}(-r/\lambda_D)$, where $\lambda_D$ is the screening length \cite {Konopka:00:Yukawa}. The screening parameter $\kappa=\Delta/\lambda_D$ was around unity in our experiment \cite{footnote1}. The particles were held together by electrostatic fields naturally present in plasma. Since the vertical confinement of particles is much stronger than their horizontal confinement \cite{Couedel:11}, they remained in a single layer at all times. We verified this by a side-view camera.

\begin{figure}
\centering
\includegraphics[width=65mm]{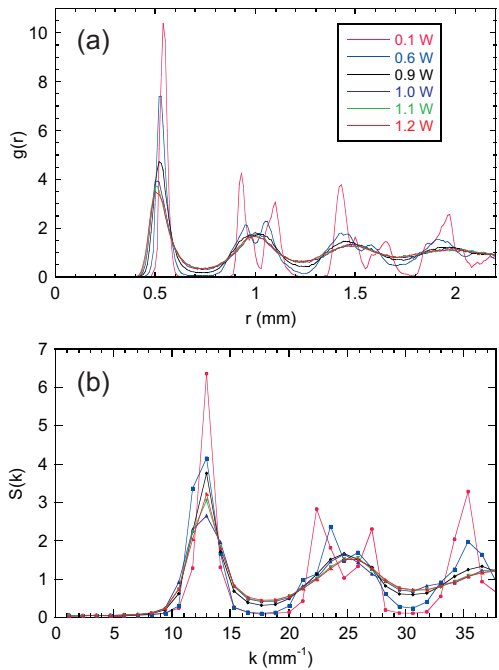}
\caption {\label {gr_Sk} Angle-averaged (a) pair correlation function $g(r)$ and (b) static structure factor $S(k)$ of the complex plasma under different levels of shear stress as controlled by the manipulation laser power $P_{\rm laser}$ (indicated in the inset). $S(k)$ was calculated (in this figure only) in a central square of twice the width of the rectangular region of interest, see Fig.~\ref{exp_layout}.}
\end{figure}

\begin{figure}
\centering
\includegraphics[width=65mm]{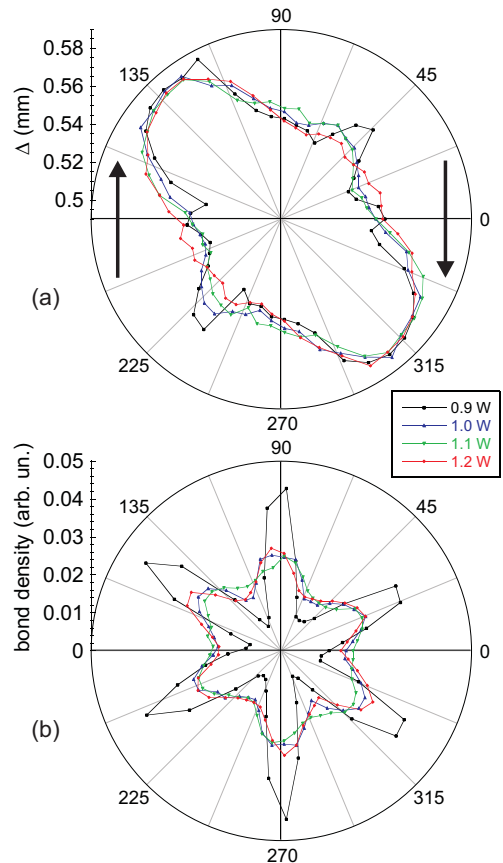}
\caption {\label {Delta_phi} (a) Interparticle distance $\Delta$ and (b) bond density as functions of azimuthal angle $\phi$ in liquid complex plasma at different levels of shear stress (controlled by $P_{\rm laser}$, see inset). The flow direction is indicated by the arrows in (a). The liquid's microstructure is anisotropic, with the compressional and extensional axes at around $\pm 45^{\circ}$ to the flow direction. The $60^{\circ}$ periodicity of the bond density in (b) indicates that the liquid complex plasma retains to some extent the original hexagonal structure of the plasma crystal (see also Fig.~\ref{S_k}). The curves for unmelted crystal ($P_{\rm laser}<0.9$~W) are discontinuous and are not shown here.
}
\end{figure}

Under our experimental conditions, the particle suspension self-organized in a highly ordered triangular lattice. We used the method of Ref.~\cite{Nosenko:04PRL_visc} to create a shear flow in the 2D complex plasma as shown in Fig.~\ref{exp_layout}. Two oppositely directed laser beams were focused down to a fraction of the interparticle spacing and they were rapidly ($\simeq 300$~Hz) scanned to draw rectangular stripes on the suspension. The particles reacted to the averaged radiation pressure. Shear stress was created in the gap between the laser-illuminated stripes, its magnitude was controlled by varying the output laser power.

To facilitate the emergence of shear flow and subsequent analysis, one of the closely packed rows of the triangular lattice was always oriented along the laser beams, before the laser was switched on. Unless otherwise stated, data analysis was performed at the stage of steady-state shear flow (after a waiting time of $3.3$~s) in a region of interest indicated by a dashed-line rectangle in Fig.~\ref{exp_layout}. All values reported were averaged over another $3.3$~s.

The transverse profiles of particle velocity in the steady-state shear flow, $v_y(x)$, were close to linear, see inset in Fig.~\ref{exp_layout}. Therefore, we assumed that the shear strain rate $\dot{\gamma}$ was uniform over the flow and calculated it as the slope of the linear fit of $v_y(x)$. In our system, the profiles of $v_y(x)$ should in fact be exponential (due to the presence of neutral gas drag force on particles), with the spatial scale defined by the (square root of) ratio of kinematic viscosity to the gas drag rate, see Ref.~\cite{Nosenko:04PRL_visc}. In the present experiment, however, the width of the flow was smaller (roughly two times) than in Ref.~\cite{Nosenko:04PRL_visc}, so the neutral gas friction played practically no role here. The transverse profiles of particle kinetic temperature $T$ were relatively flat within the region of interest shown in Fig.~\ref{exp_layout}, with a variation of $8\%-30\%$ depending on the manipulation laser power. In these conditions, the diffusive term in the heat balance for particles was at least one order of magnitude smaller than the friction term, so that the heat balance reduced to the following equation: $\nu nk_B(T-T_0)=\eta\dot{\gamma}^2/4$, where $n$, $\eta$ and $T_0$ are, respectively, areal number density, shear viscosity, and background temperature of the particle suspension \cite{Nosenko:PRL08}. Experimentally measured temperature scaled roughly as $T\simeq\mathcal{A}\dot{\gamma}$, where $\mathcal{A}=5.7\times10^4~{\rm Ks}$. This is consistent with the $\eta\propto T^{-1}$ scaling in the above equation, in the relevant temperature range \cite{Hamaguchi:02}.

We begin with standard structure analysis, i.e., calculating the pair correlation function $g(r)$ and static structure factor $S(k)$ of complex plasma (averaged over azimuthal angle $\phi$), shown in Figs.~\ref{gr_Sk}(a) and (b), respectively. Both indicate that the complex plasma is in a crystalline state for the laser power $P_{\rm laser}<0.9$~W and becomes liquid for $P_{\rm laser}\geq0.9$~W, as evidenced by a sudden decrease in the height of the first peak of $g(r)$ and $S(k)$ and disappearance of the splitting in their second peak. Although further subtler changes in both functions occur with increasing levels of shear strain rate in the liquid state, these are less prominent and difficult to decipher.

To get a greater insight on the liquid microstructure, we propose to calculate the mean interparticle spacing $\Delta$ as a function of the azimuthal angle $\phi$. This is equivalent to evaluating the first peak in the angle-resolved pair correlation function $g({\bf r})$, yet provides a very local measure and is also easier to calculate. Experimentally observed $\Delta(\phi)$ is shown in Fig.~\ref{Delta_phi}(a). A complementary measure is the angular bond density, i.e., the probability to find a near neighbor at a certain angle $\phi$ from a given particle, see Fig.~\ref{Delta_phi}(b).

The microstructure of sheared liquid is clearly anisotropic: The interparticle spacing is larger in a certain direction (extensional axis at $\phi\simeq -45^{\circ}$) and smaller in the orthogonal direction (compressional axis at $\phi\simeq 45^{\circ}$). This constitutes a departure from isotropic distribution in a quiescent liquid.

The distorted pair correlation function is given by
\begin{equation}\label{gr}
g({\bf r})=g_0(r)[1+\tau\dot{\gamma}(xy/r^2)f(r)],
\end{equation}
where $g_0(r)$ is for liquid in equilibrium, $\dot{\gamma}$ is shear rate, $\tau$ relaxation time, and the function $f(r)$ is determined by the particular form of the pair interaction potential for particles. The term proportional to $xy/r^2$ introduces a ${\rm sin}(2\phi)$ component in the angular dependence of $g({\bf r})$. Similarly, for the distorted static structure factor one has $S({\bf k})=S_0(k)[1+\tau\dot{\gamma}(k_xk_y/k^2)\mathcal{F}(k)$]. We note that these expansions for ``strained'' $g({\bf r})$ and $S({\bf k})$ are valid for small $\tau\dot{\gamma}$.

\begin{figure}
\centering
\includegraphics[width=\columnwidth]{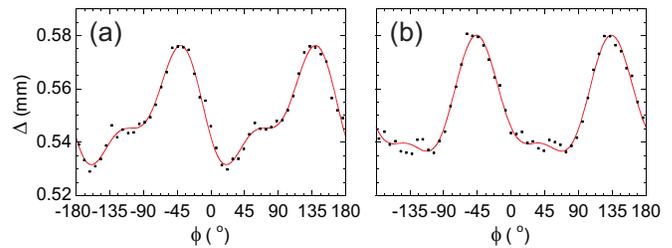}
\caption {\label {Delta_phi_fits} Fits with Eq.~(\ref{fit}) of experimental $\Delta(\phi)$ for (a) $P_{\rm laser}=1.05$~W and (b) $1.2$~W.}
\end{figure}

In our experiment, the biggest contribution to the oscillatory part of $\Delta(\phi)$ is given by ${\rm sin}(2\phi)$, with a coefficient $A_2\simeq0.035$. This term, however, does not explain all features of the observed $\Delta(\phi)$ and the inclusion of higher harmonic ${\rm sin}(4\phi)$ is necessary with $A_4\simeq0.015$. The fits of experimentally measured interparticle distance with the following ansatz:
\begin{equation}\label{fit}
\Delta(\phi)=\Delta_0[1+A_2{\rm sin}(2\phi+\delta_2)+A_4{\rm sin}(4\phi+\delta_4)]
\end{equation}
are shown in Fig.~\ref{Delta_phi_fits}.

The observed anisotropy of sheared liquid's microstructure has clear physical meaning. The flow's strain rate (normalized) is given by the degree of anisotropy, i.e., $\tau\dot{\gamma}\simeq2A_2$. The deviation of the directions of compressional and extensional axes from $\pm 45^{\circ}$ gives the extent of the non-Newtonian behavior of liquid \cite{Hanley:83}. The significance of $A_4$ is less clear, it may reflect some kind of nonlinearity.

\begin{figure}
\centering
\includegraphics[width=75mm]{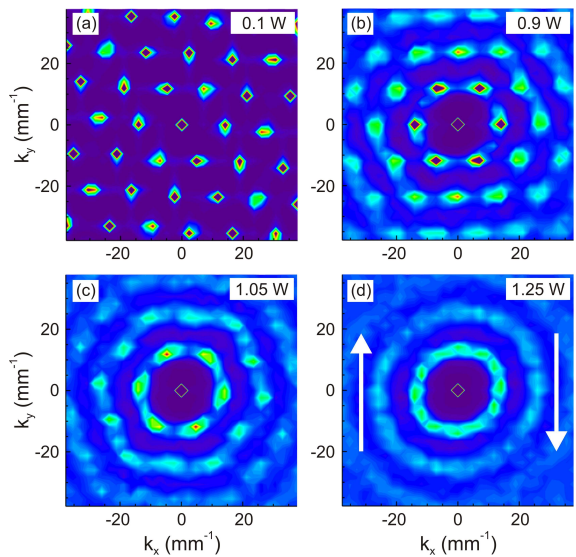}
\caption {\label {S_k} Static structure factor $S({\bf k})$ of the 2D complex plasma under different levels of shear stress (controlled by $P_{\rm laser}$, as indicated). The arrows in (d) indicate schematically the direction of the flow. Color coding from dark blue to red corresponds to the range of $0-10$ in (a) and $0-5$ in (b),(c), and (d).}
\end{figure}

Another way of evaluating the microstructure of a sheared liquid is to measure its static structure factor $S({\bf k})$ with angular resolution. For regular liquids, this is usually deduced from scattering experiments; here, we calculate $S({\bf k})$ directly from particle positions \cite{footnote2}. It is shown in Fig.~\ref{S_k} for different levels of applied shear stress. At very low stress, $S({\bf k})$ retains the structure of sharp peaks characteristic of an undisturbed triangular lattice, as shown in Fig.~\ref{S_k}(a). The peaks become blurred at higher stress as the lattice deforms and melts and eventually they coalesce into rings in fully developed shear flow, Figs.~\ref{S_k}(c),(d). The rings are distorted from the circular shape pertinent to quiescent liquids: They acquire certain ellipticity. This is particularly clearly seen in the first Debye-Scherrer ring. The ratio of its major and minor axes is given by $1+\epsilon$, where $\epsilon=\tau\dot{\gamma}$ \cite{Clark:80,Hanley:83}. This allows one to derive $\tau\dot{\gamma}$ from shear-distorted $S({\bf k})$.

\begin{figure}
\centering
\includegraphics[width=75mm]{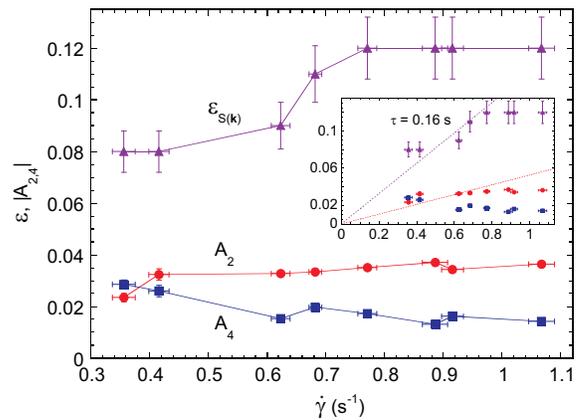}
\caption {\label {fits} Ellipticity $\epsilon$ of the liquid static structure factor $S({\bf k})$ (triangles) and relative amplitudes $A_2$ (circles) and $A_4$ (squares) in Eq.~(\ref{fit}), for different levels of shear strain rate. The inset shows linear fits (forced to $0$) of $\epsilon$ and $A_2$ for $\dot{\gamma}<0.8~{\rm s}^{-1}$.}
\end{figure}

Figure \ref{fits} shows experimentally measured $\epsilon$ and $A_{2,4}$ as functions of shear rate $\dot{\gamma}$. There is a trend for $\epsilon$ and $A_2$ to increase and for $A_4$ to diminish for larger $\dot{\gamma}$. We note that $\epsilon$ is larger than $2A_2$ by a factor of $1.2-1.6$. The uncertainty of $\epsilon$ is relatively large, since $\epsilon$ was directly measured from figures like Fig.~\ref{S_k} (not from a fit).

The resulting values of $\tau\dot{\gamma}$ depend only weakly on $\dot{\gamma}$ and saturate, for larger $\dot{\gamma}$, at about $0.11$ and $0.07$ for the $S({\bf k})$ and $\Delta(\phi)$ data, respectively. The relaxation time $\tau$ is therefore not constant but scales roughly as $\tau\propto\dot{\gamma}^{-1}$. (The inset in Fig.~\ref{fits} shows that fits with $\tau={\rm const}$ do not work well.) Recalling that the particle kinetic temperature $T\simeq\mathcal{A}\dot{\gamma}$, we then find that $\tau\simeq0.1\mathcal{A}T^{-1}$. In the range of $T=2\times10^4-6\times10^4$~K achieved in the present experiment, $\tau=0.1-0.3$~s. The temperature scaling of relaxation time hints at the particle diffusion as the process that defines it. We estimate the diffusion time in our system as $\tau_{\rm diff}=\Delta^2/D$, where $D$ is the self-diffusion coefficient. Because of the lack of systematic data on diffusion in two-dimensional complex plasmas, we use the empirical approximation for $D$ in three-dimensional equilibrium Yukawa systems proposed in Ref.~\cite{Ohta:00}: $D\simeq0.01\omega_{\rm E}\Delta^2(T/T_m-1)$, where $\omega_{\rm E}$ is the Einstein frequency \cite{Knapek:07} and $T_m$ is the (equilibrium) melting temperature that should be regarded as a dimensional parameter in our case. Since the measurements of microstructure were performed well in the liquid state, we assume that $T\gg T_m$ and obtain $\omega_{\rm E}\tau_{\rm diff}\simeq10^2T_m/T$, which is consistent with the experimentally observed temperature scaling of relaxation time. Quantitative agreement with experiment is achieved at $T_m\simeq10^{-3}\mathcal{A}\omega_{\rm E}=3.2\times10^3$~K; this is much lower than the experimental range of temperature.

To summarize, we studied for the first time the microstructure of a 2D liquid complex plasma subjected to shear flow. To evaluate the microstructure, we used two measures: the mean interparticle distance as a function of azimuthal angle $\Delta(\phi)$ and the static structure factor $S({\bf k})$. Both methods reveal anisotropic microstructure, with compressional and extensional axes at around $\pm 45^{\circ}$ to the flow direction. The ellipticity of $\Delta(\phi)$ or $S(\bf{k})$ yields the (normalized) shear rate of the flow. The two measures give comparable results.

We thank J\"{u}rgen Horbach for valuable discussions. The research has received funding from the European Research Council under the European Union's Seventh Framework Programme (FP7/2007-2013) / ERC Grant agreement 267499.

\end{document}